\documentclass[12pt]{article}
\usepackage{graphicx}
\usepackage{natbib}
\usepackage{txfonts}
\bibpunct{(}{)}{;}{a}{}{,}

\newcommand{\kms}{\,km\,s$^{-1}$}

\newcommand{\ro}{\,$R_{\odot}$}

\date{}
\begin{document}
\markboth{Skopal}{The light curves of symbiotic stars}
%
%
\title{How to understand the light curves of symbiotic stars}
%
\author{Augustin Skopal \\
        Astronomical Institute, Slovak Academy of Sciences,\\
        SK-059\,60 Tatransk\'{a} Lomnica, Slovakia 
        (e-mail: skopal@ta3.sk)}
%
%
\maketitle
\begin{abstract}
 I introduce fundamental types of variations observed in the 
 light curves (LC) of symbiotic stars: the orbitally-related 
 wave-like modulation during quiescent phases, eclipses 
 during active phases and apparent orbital changes indicated 
 during transitions between quiescence and activity. 
 I explain their nature with the aid of the spectral energy 
 distribution (SED) of the composite spectrum of symbiotic 
 stars and their simple ionization model. 
\end{abstract}

\section{Introduction}

The symbiotic stars are understood as interacting binary 
systems comprising a late-type giant and a hot compact
star -- most probably a white dwarf. Their orbital periods
run usually between 1 and 3 years, but can be significantly 
larger.
The mass loss from the giant represents the primary condition 
for appearance of the symbiotic phenomenon. A part of the 
material lost by the giant is transferred to the compact 
companion via accretion from the stellar wind. This process 
makes the accretor very hot 
  ($T_{\rm h} \sim 10^5$\,K)
and luminous
  ($L_{\rm h}\sim 10-10^4\,L_{\odot}$),
and thus to be capable of ionizing a fraction of the neutral 
wind from the giant, giving rise to nebular emission. As a result 
the spectrum of symbiotic stars consists of three basic components 
of radiation -- two stellar (from the binary components) and one 
nebular, emitting by the ionized winds of both the stars. 

If the processes of the mass-loss, accretion and ionization 
are in a mutual equilibrium, then symbiotic system releases 
its energy approximately at a constant rate and SED. 
This stage is known as the {\em quiescent phase}. Once this 
equilibrium is disturbed, symbiotic system changes
its radiation significantly, at least in its SED, 
which leads to a brightening in the optical by a few 
magnitudes. We call this the {\em active phase}. 

The presence of physically different sources of radiation 
in the system which differ extremely in temperatures, and 
also their nature (stellar and nebular component), 
produce a complex composite spectrum. The resulting spectrum 
thus depends on the wavelength, the activity of the system 
and also the projection of these regions into the line of 
sight, i.e., on the orbital phase of the binary. In addition, 
the composite spectrum of individual objects is also a function 
of their physical and orbital parameters. 
Throughout the optical, the light contributions from these 
sources rival each other, producing a spectrum whose color 
indices differ significantly from those of standard stars. 

Therefore the LCs of symbiotic stars have a complex profile, 
often having an unexpected variation. Generally, the most 
pronounced changes are observed at the short-wavelength domain 
of the visual region -- namely within the photometric $U$ filter. 
In this passband the dominant light contribution usually comes 
from the nebula, which responds most sensitively to the variation 
of the energy production of the symbiotic system. The large variety 
of changes recorded in LCs of symbiotic stars is very broad, 
and many variations are not quite understood yet. 
It is, however, clear that they are related to those from 
X-rays to radio wavelengths. From this point of view, 
photometric monitoring is important to complement other 
multifrequency observations and thus to help in understanding 
the responsible physical processes. This aspect was recently 
highlihted by \cite{sok03} and demonstrated for the case 
of the symbiotic prototype Z\,And by \cite{sok+06}. 

In this contribution I discuss just the fundamental types of 
variations in the LCs of symbiotic stars that reflect most 
closely their nature -- the orbitally-related wave-like 
variation, eclipses and apparent changes of the orbital period. 
To understand these types of variability, I compare the 
multicolor LCs with the disentangled composite spectrum 
in the visual domain and consider basic ionization structure 
of symbiotic stars. First, I introduce some examples 
of their LCs. 
%
%
\begin{figure}[!t]
\centering
\begin{center}
\resizebox{15cm}{!}{\includegraphics[angle=-90]{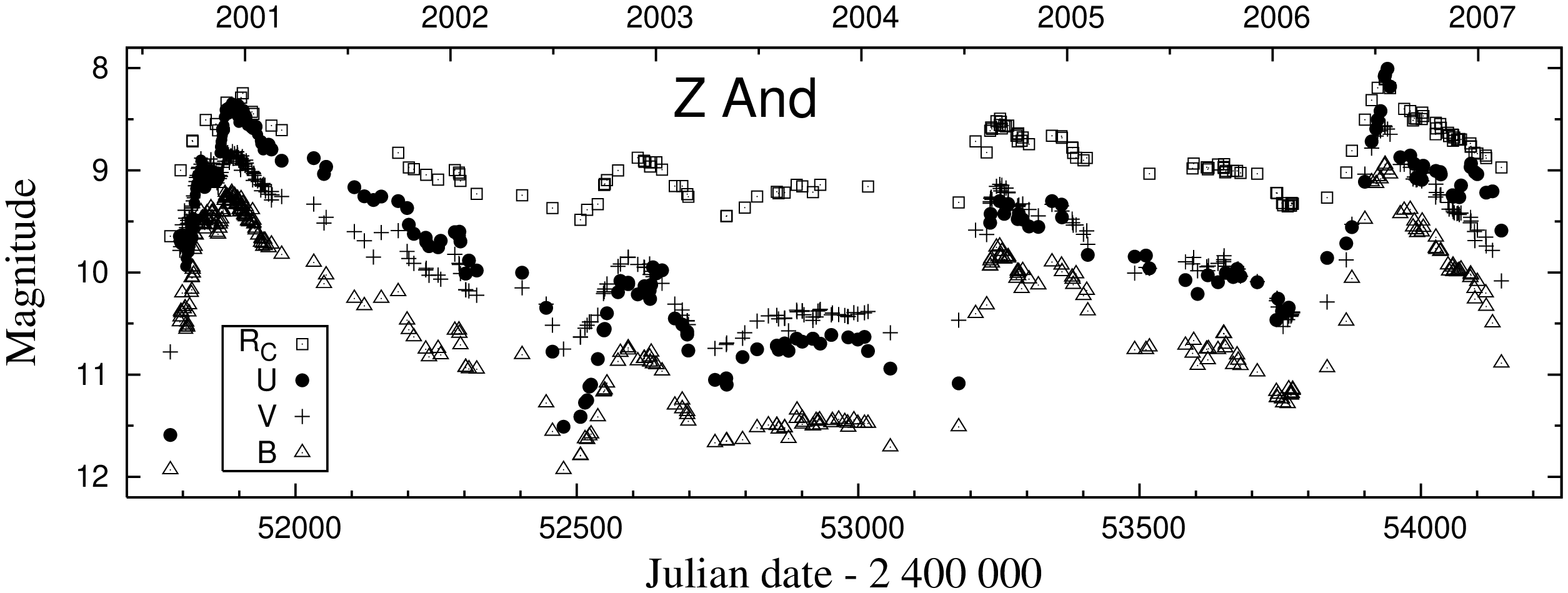}}
\resizebox{15cm}{!}{\includegraphics[angle=-90]{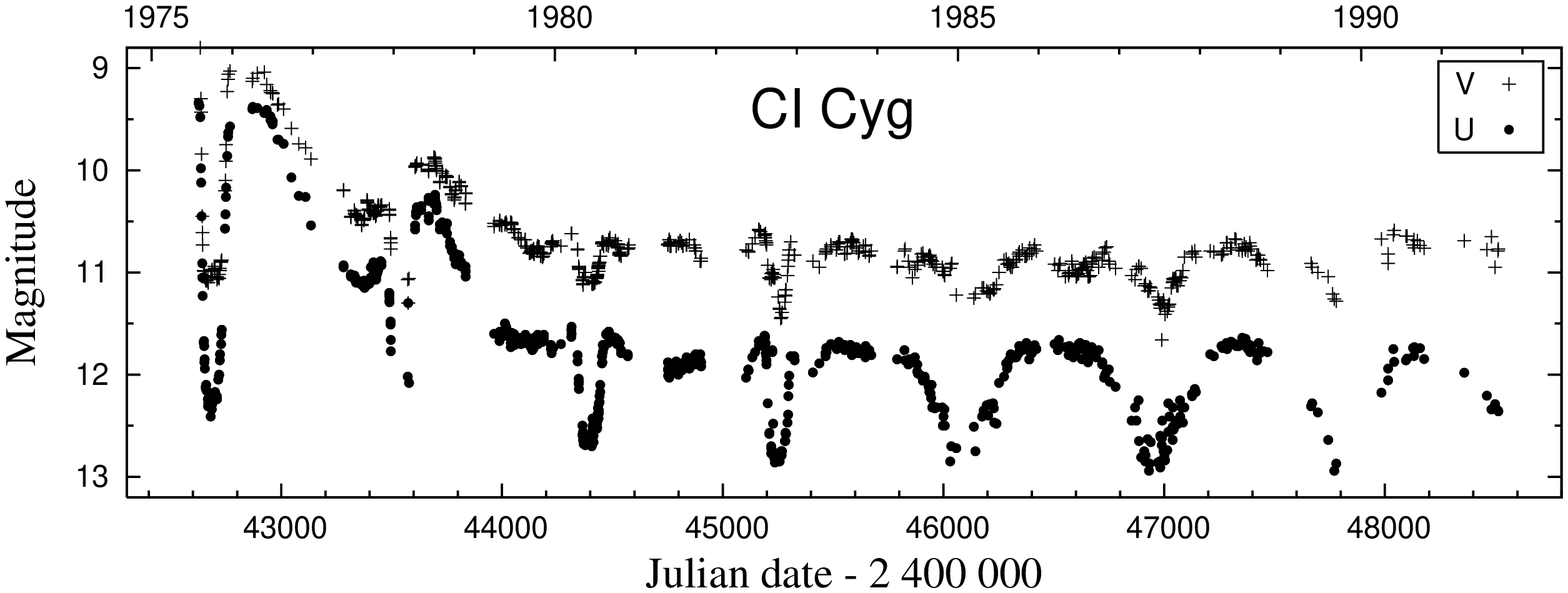}}
\resizebox{15cm}{!}{\includegraphics[angle=-90]{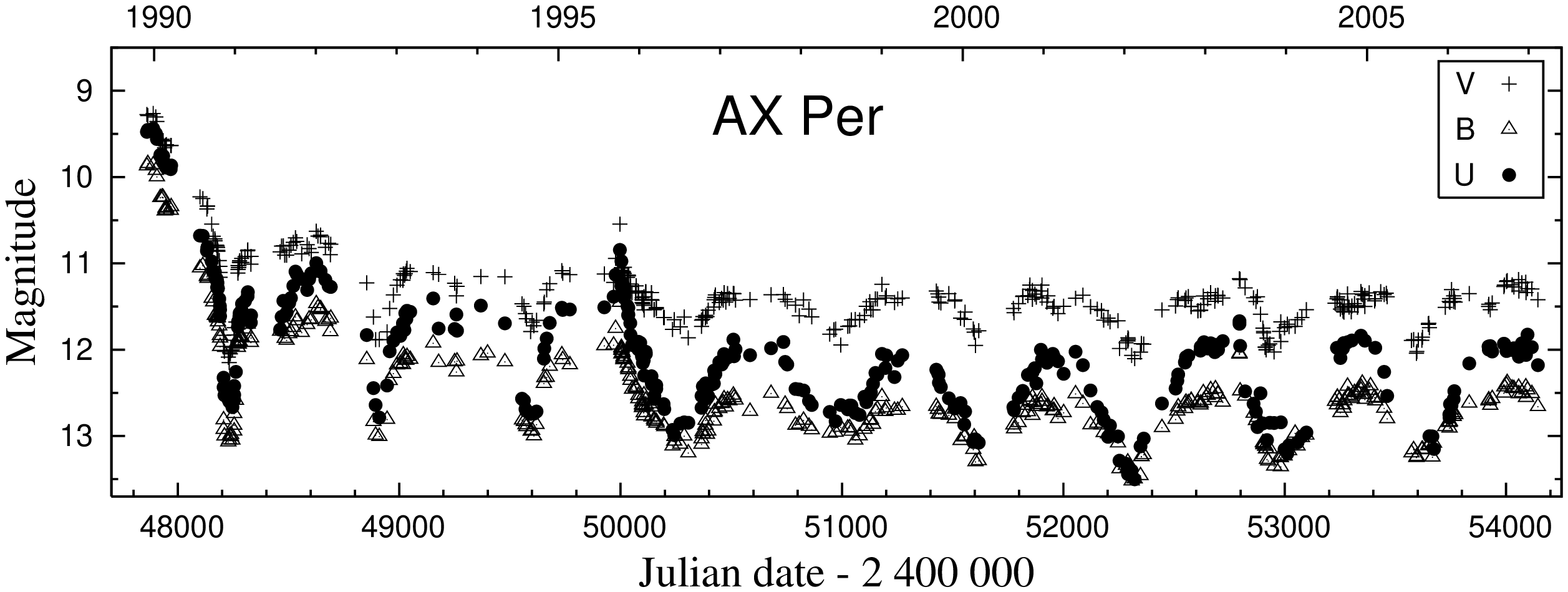}}
\caption[]{
Top: 
The $UBVR_{\rm C}$ LCs of Z\,And covering two major 
eruptions that peaked in 2000 December and 2006 July 
\citep[from][]{sk+07}.
Middle: 
Example of $UV$ LCs of CI\,Cyg from its 1975 outburst 
with narrow minima -- eclipses. From about 1984 eclipses 
transferred into wave-like variation signaling thus quiescent 
phase. The data are from \cite{bel92}. 
Bottom: The $UBV$ LCs of AX\,Per covering a part of its 
1989-94 active phase and the following quiescence from 1995 
\citep[see][]{sk+01}. 
          }
\end{center}
\end{figure}
%
%
\section{Examples of LCs of well studied symbiotic binaries}

\subsection{Z\,Andromedae}

Z\,And is considered a prototype of the class of symbiotic
stars. The binary comprises a late-type, M4.5\,III, giant
and a white dwarf accreting from the giant's wind on the 758-day
orbit \citep[e.g.][]{nv89}. More than 100 years of monitoring 
Z\,And has shown the eruptive character of its LC. It displays
several active phases, during which fluctuations range in 
amplitude from a few tenths of a magnitude to about 3 magnitudes 
\citep[][]{fl94}. Figure~1 (top panel) displays its recent activity 
from 2000 autumn covering the optical maxima in 2000 December, 
2004 September, and 2006 July. 

\subsection{BF\,Cygni}

BF\,Cyg is an eclipsing symbiotic binary with an orbital 
period of 757.2\,d \citep[][]{fekel01}. The system consists 
of a late-type cool component classified as a normal M5\,III 
giant \citep[][]{ms99} and a hot, luminous compact object 
\citep[][]{mkm89}. Its historical LC shows three basic types 
of active phases -- nova-like and Z\,And type of outburst 
and short-term flares \citep[][]{sk+97}. 
Figure~2 (top panel) shows its LC from 1985 covering the recent 
1989-outburst with an eclipse effect and wave-like variation 
during the following quiescent phase. 

\subsection{CI\,Cygni}

CI\,Cyg is also eclipsing symbiotic binary with the orbital 
period of 855.25 days \citep[][]{bel79,bel84}. Its cool component 
was recently classified as a M5.5\,III giant \citep[][]{ms99}. 
A detailed study of this system was made by \cite{ken+91}. 
The last major active phase of CI\,Cyg began in 1975 \citep{bel76}. 
During the first four cycles from the maximum, narrow
minima indicating eclipses developed in the LC, a typical feature
of active phases of symbiotic stars having a high orbital
inclination. From 1985 the minima profile became very broad 
indicating a quiescent phase (Fig.~1, mid panel). From 2006 May 
CI\,Cyg entered its new active phase \citep[][]{sk+07}. 

\subsection{V1329\,Cygni}

The symbiotic phenomenon of V1329\,Cyg developed during 
its nova-like eruption in 1964. Prior to this outburst, 
V1329\,Cyg was an inactive star of about 15th magnitude 
displaying $\sim$2\,magnitude-deep eclipses 
\citep[see Fig.~1 in][]{mu+88}. 
The post-outburst LC shows large, $\sim$1.5\,magnitude deep, 
periodic, wave-like variations connected with the binary motion. 
The IUE observations revealed the presence of a strong 
nebulosity in the near-UV spectrum (Fig.~5). 

\subsection{AG\,Draconis}

This symbiotic system belongs to the group of so-called 
yellow symbiotics, because it contains a K2\,III giant as 
a cool component \citep[][]{ms99}. There are no signs 
of eclipses either in the optical or the far-UV regions. 
\cite{ss97}, based on spectropolarimetric observations, 
derived the orbital inclination $i=60\,(\pm 8^{\circ}.2)$. 
The system undergoes occasional eruptions. The star's brightness 
abruptly increases by 1--3\,magnitudes, often showing multiple 
maxima separated approximately by 1 year 
\citep[][]{lut83, viotti+07}. 
The quiescent phase of AG\,Dra is characterized by a periodic 
wave-like variation, which is more pronounced at shorter 
wavelengths. Figure~2 (mid panel) shows a part of its recent 
LC covering both the quiescent and the active phases. 

\subsection{AX\,Persei}

AX\,Per is known as eclipsing symbiotic binary with 
an orbital period of 680 days \citep[][]{sk91}. 
The cool component of the binary is a normal giant of the 
spectral type M4.5\,III \citep[][]{ms99}. 
The historical LC of AX\,Per is characterized by long-lasting 
periods of quiescence with the superposition of a few bright 
stages \citep[see Fig.~1 of][]{sk+01}. 
Figure~1 (bottom panel) demonstrates evolution in the LC 
covering a part of its last active phase with eclipses (1990-94) 
and the transition to quiescence at 1995.8, followed by 
typical periodic waves in the star's brightness. 
%
%
\begin{figure}[!t]
\centering
\begin{center}
\resizebox{15cm}{!}{\includegraphics[angle=-90]{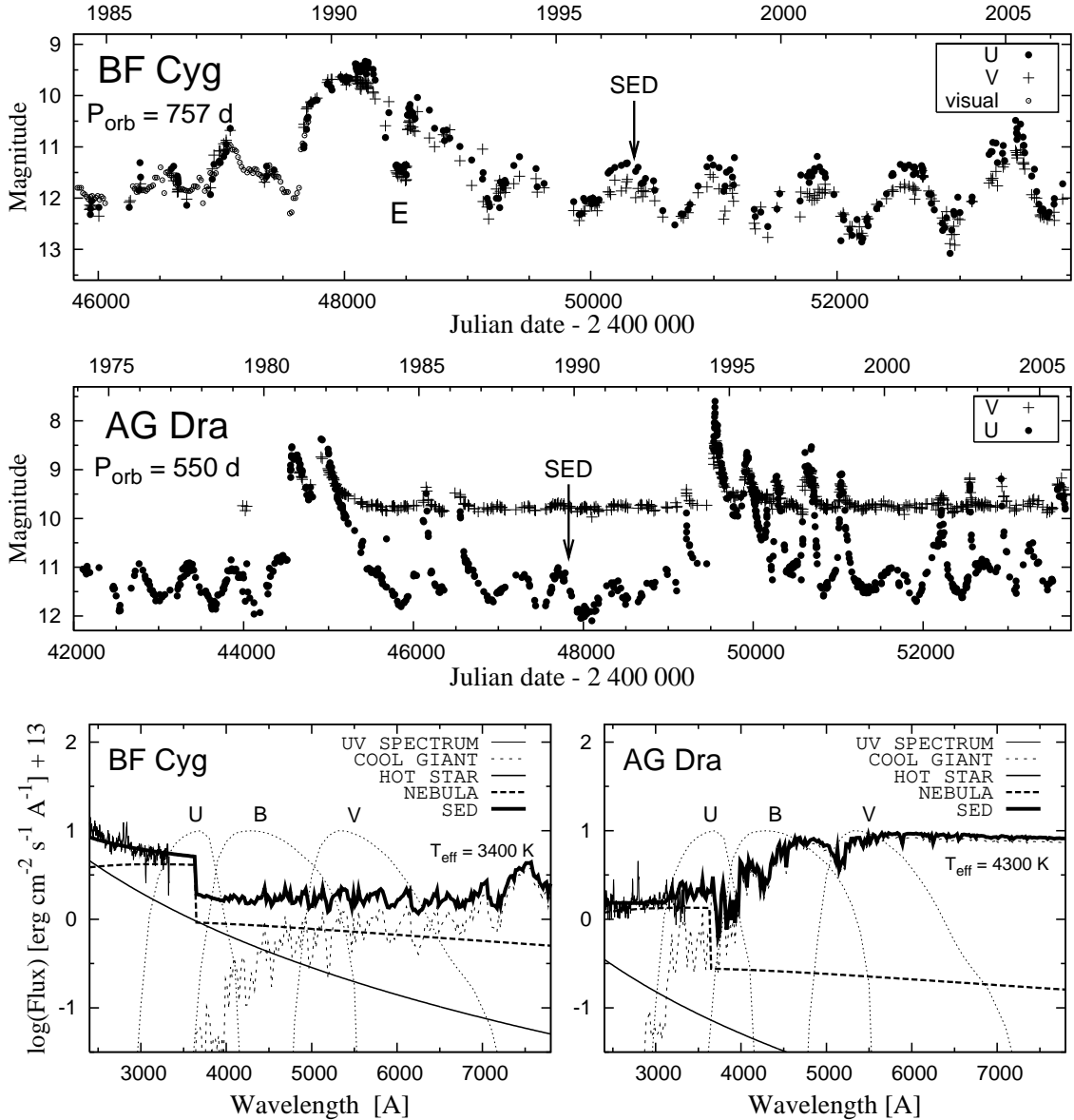}}
\caption[]{Top panels show the LCs of BF\,Cyg and AG\,Dra in 
$U$ and $V$ filters. During active phase the eclipsing system 
BF\,Cyg displays a relatively narrow minimum at the inferior 
conjunction of the giant (denoted by {\sf E}), while during 
quiescent phase its LC shows pronounced wave-like variation, 
characterized with amplitudes 
$\Delta U \approx \Delta V \sim 1.5$\,mag. 
The yellow symbiotic star AG\,Dra is not eclipsing. Amplitudes 
of its wave-like variations are smaller and depend considerably 
on the color: $\Delta U \sim 1$\,mag, $\Delta V \sim 0.1$\,mag. 
The bottom panels show the SEDs of these objects throughout 
the $UBV$ passbands, which explains the observed differences 
in the wave-like variation (Sect.~3). 
          }
\end{center}
\end{figure}
%
%
\section{Wave-like orbitally-related variation}

Wave-like, orbitally-related variability represents the most 
characteristic feature of the LCs of symbiotic stars that 
develops during their quiescent phases.
Generally, we observe a periodic, wave-like profile of LC, 
whose minima and maxima occur at or around conjunctions of 
the binary components. The inferior conjunction of the giant 
(the cool component in front of the hot star) corresponds 
to the light minimum (orbital phase $\varphi = 0$), while 
at its superior conjunction (the hot star in front) 
we observe a light maximum ($\varphi = 0.5$). This variation 
is characterized with a large magnitude difference between 
the minimum and maximum, $\Delta m \sim 1-2$\,magnitudes. 
This 'amplitude' is always larger in the blue part of the 
spectrum than in the red one, 
i.e. $\Delta U > \Delta B > \Delta V$. 
This relationship can be understood with the aid of the SED 
throughout the $UBV$ region. 

Figure~2 shows examples of this type of light variations 
for BF\,Cyg (it contains a red M5 giant with the effective 
temperature $T_{\rm eff} \sim 3\,400$\,K) and AG\,Dra 
(yellow K2 giant with $T_{\rm eff} \sim 4\,300$\,K) 
with their SEDs covering the optical domain. This spectral 
region is dominated by the radiation from the nebula and 
the giant. The latter does not depend on the orbital phase 
and strengthenes considerably towards the longer wavelengths, 
while the nebular radiation has the opposite behavior (it 
dominates the $U$ passband and is fainter in $V$) and it is 
the source of the orbitally-related variation (Sect.~5.2). 
Therefore the $\Delta m$ amplitudes are declining to longer 
wavelengths, where the nebular emission is superposed 
with the increasing light from the giant, which does not 
vary with the orbital motion. 
   In other words, the observed amplitude of the wave-like 
variation is proportional to the ratio of fluxes from the 
nebula and the giant, which is a function of the 
wavelength -- flux from the giant/nebula increases/decreases 
with increasing lambda. 

In the cases of the so-called yellow symbiotic stars (they 
contain a giant of the spectral type K to G), the giant's 
contribution into the $V$ passband is very strong, which 
produces very different $\Delta U$ and $\Delta V$ amplitudes: 
$\Delta U$/$\Delta V >> 1$. If the system contains a red 
giant, its contribution in $V$ is relatively lower in the 
total composite spectrum, which yields the ratio 
$\Delta U$/$\Delta V \ge 1$. 
In our example on Fig.~2, $\Delta U$/$\Delta V \sim 1.4$ 
for BF\,Cyg, whereas for the yellow symbiotic star AG\,Dra 
$\Delta U$/$\Delta V \sim 10$ \citep[see also Fig.~25 in][]{sk05}. 
On the other hand, a markedly different amplitude in $U$ 
(eventually in $B$) and $V$ (eventually in $R$) filters 
signals the presence of a yellow cool component in the 
symbiotic system. 

\section{Eclipses}

During active phases of systems with high orbital inclination, 
a significant change in the minima profile is observed -- 
the very broad profile becomes narrow. As the minima 
coincide with the inferior conjunction of the cool component, 
it is believed that they are caused by eclipses of the hot 
object by the cool giant. Examples of this effect are shown 
in Fig.~1 and top panels of Fig.~3 for eclipsing 
symbiotic binaries BF\,Cyg, AX\,Per and CI\,Cyg.

According to spectroscopic observations an optically thick 
shell -- a false photosphere -- is created around the hot 
active star, which redistributes a significant fraction of 
its radiation. As the characteristic temperature of the false 
photosphere \citep[$\sim 22\,000$\,K,][]{sk05} is considerably 
lower than that of the hot star during the quiescence 
($\sim 10^5$\,K), its light contribution will be shifted 
to longer wavelengths according to the Wien's displacement 
law, and thus make the visual region brighter. 
The bottom left panel of Fig.~3 shows an example of 
BF\,Cyg during its 1990 major outburst. The hot star 
pseudophotosphere radiates at the temperature 
$T_{\rm h} = 21\,500$\,K, and its contribution is above 
those from the giant and the nebula through the 
$UBV$ domain. Since the radius of the false photosphere 
is of a few solar radii, the cool giant can eclipse it 
easily for about one tenth of the orbital period 
(i.e. 2 -- 3 months) that corresponds to typical giant's 
radius of 100\ro\ and orbital periods as long as 2 -- 3 years. 
The depth of eclipses usually obeys the relation: 
$\Delta U > \Delta B > \Delta V$, because the light from 
the hot star decreases towards the red part of the spectrum, 
while that from the giant increases. However, the resulting 
eclipse depth and color indices are modulated by the presence 
of a rather strong nebula in the system, which is not subject 
to eclipses (bottom mid panel of Fig.~3). Thus, during 
totality the nebula partially fills-in the minima and, in 
combination with the radiation from the giant, produces color 
indices that differ significantly from those of a normal red 
giant. For example, we observed $U-V \sim 0$, 
+0.5 and +1.2 for BF\,Cyg, AX\,Per and CI\,Cyg, respectively, 
during their total eclipses (cf. Fig.~3 top). For a comparison, 
in a theoretical case that the nebula is not present outside 
the eclipsing giant's stellar disk, we should measure the color 
indices of a normal red ginat, 
e.g. $U-V \sim +3$\,mag \citep[e.g.][]{lee70}. 
On the other hand, knowing the spectral type of the giant 
and having measured magnitudes at totality would allow us 
to estimate parameters of the contributing nebula -- its 
emission measure and the electron temperature. 

During quiescent phase, radiation from the nebula 
dominates the optical -- its contribution to the $UBV$ 
passbands is well above those from the hot star and the giant 
(cf. Fig.~3 bottom right). The nebula represents a very 
extended source of radiation in the symbiotic system, which 
thus cannot be subject to eclipse. As a result, we instead 
observe a very broad minima; the LC waves as a function of 
the orbital phase. 
%
%
\begin{figure}[!t]
\centering
\begin{center}
\resizebox{15cm}{!}{\includegraphics[angle=-90]{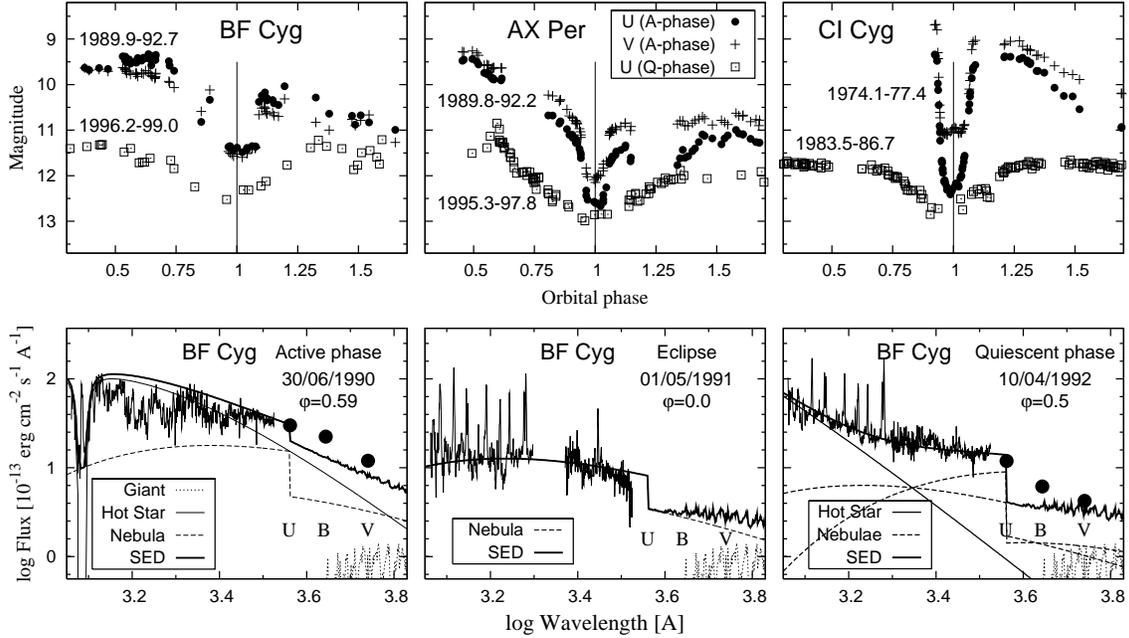}}
\caption[]{
Eclipses + wave variation and SEDs.
Bottom: Examples of the SED during the active phase (left panel), 
eclipse (middle) and quiescent phase (right) of BF\,Cyg. 
During activity contribution from a warm false photosphere 
around the hot star is larger than that from the nebula in 
the $UBV$ region. As a result we observe narrow minima -- 
eclipses -- in the LC at the inferior conjunction of the 
giant. The minima are in part filled in with rather strong 
residual light from the nebula, which is not subject to 
eclipse (mid panel). During quiescence the radiation from 
extended nebula(e) dominates the $UBV$ region (right panel), 
which causes the minima to be very broad. 
Top panels demonstrate how these SED variations are reflected 
by the LCs of eclipsing systems BF\,Cyg, AX\,Per and CI\,Cyg. 
          }
\end{center}
\end{figure}
%
%

\section{On the nature of the wave-like variations}

\subsection{Reflection effect}

Originally, \cite{boy66} and \cite{bel70} suggested a reflection 
effect as being responsible for the periodic wave-like variation 
recorded in their LC of AG\,Peg. 
In this model, the hot star irradiates and heats up the facing 
giant's hemisphere that causes variation in the star's brightness 
when viewing the binary at different orbital phases. The left 
panel in Fig.~4 illustrates the scheme of the reflection effect 
as suggested by \cite{ken86}. 
At the inferior conjunction of the giant ($\varphi \sim 0$) 
we observe a minimum of the light, and conversely, at the 
giant's superior conjunction ($\varphi \sim 0.5$) we observe 
its maximum analogous to the Moon's phases. 
This natural explanation was adopted by many authors 
\citep[e.g.][]{ken86, mun89} and it is still considered 
as a possible cause of the wave variation in LCs as a function 
of the orbital phase \citep[][]{m+js02}. 

However, the reflection effect fails to explain quantitatively 
the observed very large amplitudes of 1--2\,mag or more, because 
a normal red giant does not intercept enough radiation from 
the hot component to produce the strong emission spectrum 
and its variation. For symbiotic stars this case was 
investigated theoretically by \cite{proga96}, who found that 
the magnitude difference between the illuminated and 
non-illuminated red giant hemisphere is less than 0.3\,mag. 
Also \cite{sk01} demonstrated that the observational 
characteristics of the LCs of symbiotic binaries -- the large 
amplitude, the profile of minima and variation in their 
positions (see Sect.~5.2) -- cannot be reproduced by 
the reflection effect. 
A better agreement between the observed and calculated 
variation in both the line and the continuum spectrum 
was achieved by including the neutral wind of the giant into 
the model, which thus could intercept much larger amount of 
the hot star radiation \citep[][]{proga98}. 
It became clear that the nature of the orbitally-related 
changes in the optical/near-UV continuum should be explained 
within the ionization model of symbiotic binaries, in which 
the hot star radiation ionizes a portion of the neutral wind 
from the cool giant. 

\subsection{Ionization model and the wave-like variability}

{\em A simple model. }
The right panel of Fig.~4 shows the ionization structure 
given by the H\,{\small II}/H\,{\small I} boundaries between 
the ionized and neutral hydrogen in a symbiotic binary 
calculated for a gradual acceleration of the giant's wind 
with the terminal velocity of 20\kms\ and a steady state 
case (binary rotation and the gravitational attraction
on the wind particles were neglected). 
The model was originally outlined by \cite{stb}(hereafter STB) 
to explain the radio emission from symbiotic stars and 
elaborated later by \cite{nv87} as a new approach to symbiotic 
stars to determine their basic physical parameters 
\citep[][]{m+91, sk05}. 
The ionization boundary is a curve at which the flux of 
ionizing photons from the hot star is balanced by the 
flux of neutral particles (here we consider just hydrogen) 
from the cool star. By other words, it is defined by the 
locus of points at which ionizing photons are completely 
consumed along paths outward from the ionizing star. 
The shape of the boundary is thus given mainly by the binary
properties -- separation of the components, number of hydrogen
ionizing photons, the mass-loss rate from the giant, and 
terminal velocity of the wind particles 
\citep[see][in detail]{stb,nv87}. 
Figure~4 shows examples of three H\,{\small II}/H\,{\small I} 
boundaries: 

   (i) The case when the flux of ionizing photons exceeds 
significantly that of neutral particles corresponds to a very 
extended symbiotic nebula -- the neutral H\,{\small I} zone 
has a cone shape with the giant at its top. 

   (ii) If both the fluxes are approximately equal, dimensions 
of both the zones are comparable. 

   (iii) For a very low ionizing capability of the hot star 
the H\,{\small II} zone can be closed around the hot star. 

The SEDs in Figs.~2 and 3 demonstrate that the nebula represents 
a significant source of the light in the visual region, mainly 
during quiescent phases. Figure~4 then suggests that this source 
of radiation is physically displaced from the giant that 
excludes directly the reflection effect to be responsible for 
the orbitally related wave-like variation in LCs. Therefore 
the principal question is how and why the symbiotic nebula 
can affect the observed light to explain this type of 
variability. In the following sections I will try to answer 
these questions. 
%
%
\begin{figure}[!t]
\centering
\begin{center}
\resizebox{12cm}{!}{\includegraphics{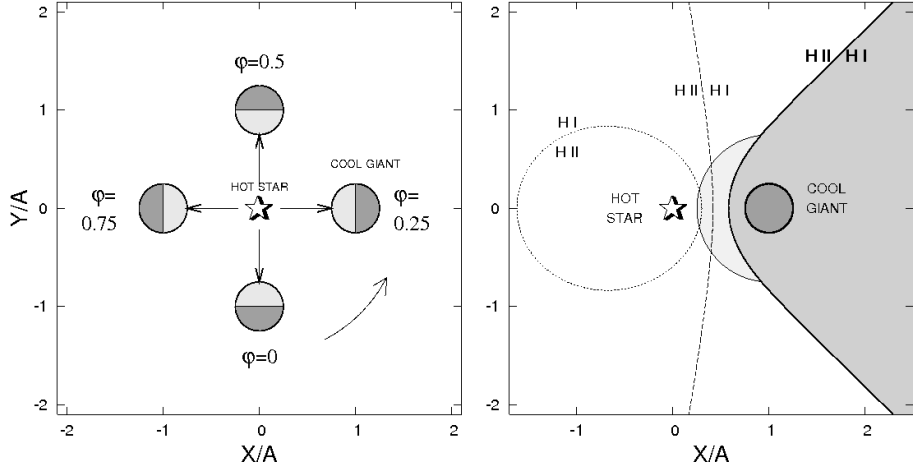}}
\caption[]{Left: Schematic representation of the reflection effect. 
Within this model the wave-like orbitally-related variation in LCs 
of symbiotic stars results from different visibility of 
the illuminated giant's hemisphere (the lighter one facing 
the hot star) at different phases. This model, however, neglects 
the effect of ionization of the neutral wind from the giant. 
Right: The STB ionization structure of the hydrogen in symbiotic 
binary. The boundary between the ionized and neutral hydrogen 
(H\,{\small II}/H\,{\small I}) for a strong (heavy solid line), 
moderate (dashed line) and faint (dotted line) source of the 
ionizing photons (i.e. the hot star). Within this model the 
wave-like variability results from a different projection of 
the optically thick portion of the ionized zone (the light 
gray part of the H\,{\small II} zone) into the line of sight 
(Sect.~5.2). 
          }
\end{center}
\end{figure}
%
%
%
%
\begin{figure}[!t]
\centering
\begin{center}
\resizebox{10cm}{!}{\includegraphics{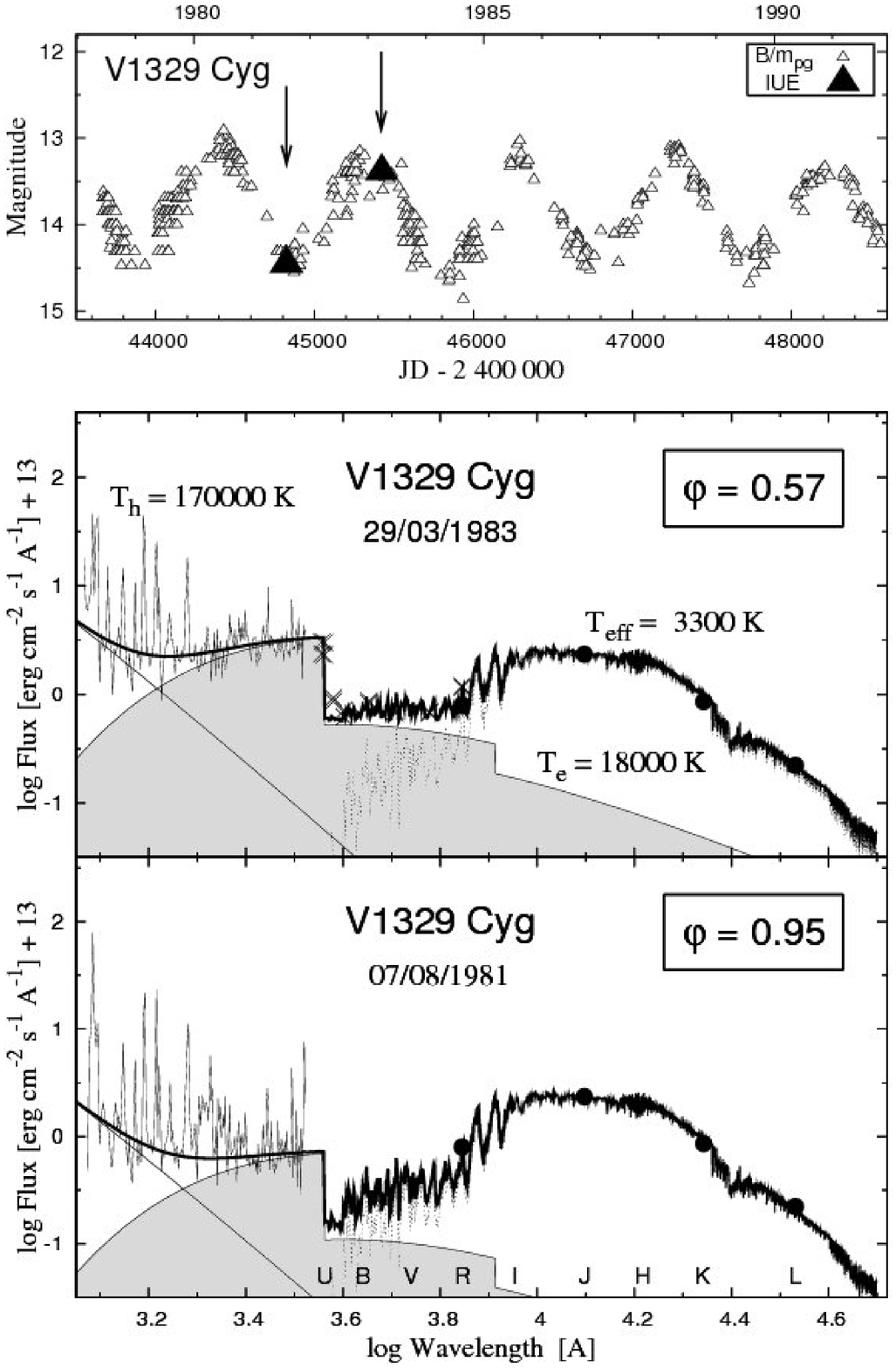}}
\caption[]{
Example of the wave-like orbitally-related variation in 
the LC of V1329\,Cyg during quiescent phase (top). It is 
caused by variation in the quantity of the nebular radiation 
observed at different orbital phases (bottom panels). 
Magnitudes derived from the IUE spectra agree perfectly with 
those obtained photometrically (filled triangles in the top 
panel). 
This result thus demonstrates that the periodic variation 
in the nebular radiation is responsible for that observed 
in the LCs. 
          }
\end{center}
\end{figure}
%
%
%

\noindent
{\em Variation in the nebular emission and LCs. }
\cite{sk01} found a relationship between the wave-like 
variation in LCs and the radiation from the symbiotic nebula. 
Both dependencies are of the same type. We observe 
a maximum/minimum of the nebular emission around the 
conjunctions of the binary components, similar to the 
periodic, wave-like variation of the photometric magnitudes. 
Figure~5 demonstrates this case for V1329\,Cyg (Sect.~2.4). 
Maximum/minimum of the nebular emission at 
$\varphi$ = 0.57/0.95 (bottom panels) corresponds to 
the maximum/minimum in the LC (top). 

Thus the orbitally related variation in the nebular 
component of radiation causes that which is observed 
in LCs. This relationship can be verified by converting 
the observed amount of the nebular radiation, usually 
characterized by emission measure, $EM$, to the scale 
of magnitudes. 
\footnote{
  The flux produced by the nebula of a volume $V$ with 
  concentrations of ions (protons), $n_{+}$, and electrons, 
  $n_{\rm e}$, largely depends on the number of hydrogen 
  recombinations, and is proportional to 
  $\int \!n_{+}n_{\rm e}\,dV$ -- the so-called emission 
  measure} 
Under the assumption that the light from the nebula dominates 
the considered passband, which is usually satisfied for $U$, 
the stellar magnitude, for example, $m_{\rm U}$, can be 
expressed as 
\begin{equation}
  m_{\rm U} = -2.5\log(EM) + C_{\rm U},
\end{equation}
where the constant $C_{\rm U}$ depends on the volume emission 
coefficient and the contribution of the zero magnitude 
star in $U$ \citep[see][ in detail]{sk01}. Figure~5 shows 
a very good agreement between the $B$-magnitudes determined  
according to Eq.~(1) and those obtained by standard 
photometric measurements. 
This result thus confirms the unambiguous connection between 
variations in the nebular emission and the photometric 
measurements. 
%

\noindent
{\em Why does the emission measure vary? }
It is simple to imagine that the orbitally-related variation 
in the nebular emission are only apparent. This implies that 
a fraction of the nebular medium has to be partially optically 
thick to produce different contributions of its total emission 
into the line of sight at different orbital phases. 
Within the STB model the opacity, $\kappa$, of the ionized 
emission medium decreases with the distance $r$ from the giant, 
since $\kappa \propto density(r) \propto r^{-2}$. This 
implies that the parts of the nebula closest to the boundary 
between the stars will be the most opaque. 

In the case of the extensive emission zone the optically thick 
portion of the H\,{\small II} region has the geometry of 
a canopy located on the boundary around the binary axis 
(Fig.~4 right). 
Such a shape will attenuate most of the radiation at 
orbital phase 0 (i.e. the relatively largest part of the 
optically thin nebula will be obscured by it), while 
at phase 0.5, we will observe a maximum light from the 
nebula in agreement with the variation in $EM$ and the LCs. 
In this case the LC profile will be a simple sinusoid. 

In the case of an oval shape of the H\,{\small II} zone 
(Fig.~4, the dotted curve) its total emission will be attenuated 
more at positions of the binary component's conjunctions (the 
orbital phases $\varphi$ = 0 and 0.5) than at positions of 
$\varphi$ = 0.25 and 0.75, when viewing the binary from its 
sides. Such apparent variation in the $EM$ can produce 
the primary, but also a secondary minimum in the LC. 
The secondary minima of this nature are well demonstrated 
by the $U$-LC of EG\,And \citep[see Fig.~2 in][]{sk05}. 

The above described approximation of the nebula shaping in 
a symbiotic system allow us to explain qualitatively just the 
most pronounced features of the LCs. A more accurate ionization 
structure, including the effect of the binary rotation and 
the gravitational attraction on the wind particles, has not 
been investigated yet. 
Nevertheless, an asymmetry of the nebula with respect to 
the binary axis (i.e. the line connecting the stars) can be 
indicated observationally. The recently discovered effect 
of apparent changes in the orbital period gives evidence of 
this possibility. 

\section{Apparent changes in the orbital period}

This effect is connected with transitions between active 
and quiescent phases of a symbiotic system. Aside from the 
significant change of the minima profile during these periods 
(Sect.~4, Fig.~3), a systematic variation in the minima 
position, i.e. the effect of apparent orbital changes, was 
revealed \citep{sk98}. 

\subsection{Systematic variation in the $O-C$ residuals}

Here I will demonstrate this effect on the historical LC 
of the eclipsing symbiotic system BF\,Cyg (Fig.~6). 
First, we determine positions of the observed ('O') minima in 
its LC and calculate those ('C') using a reference 
ephemeris. Then we construct the so-called $O-C$ diagram 
(i.e. the residuals between the observed and calculated timing 
of the minima). In our example of BF\,Cyg the $O-C$ diagram 
was constructed using the reference ephemeris given by all 
the primary minima measured by \cite{sk98} 
\begin{equation}
 JD_{\rm Min} = 2\,411\,268.6 + 757.3(\pm 0.6)\times E,
\end{equation}
which is identical (within uncertainties) with the spectroscopic
ephemeris of \cite{fekel01}. 
A systematic variation in the $O-C$ residuals is clearly seen. 
This behavior was already noted by \cite{jac}. The gradual 
increase of the $O-C$ values before the 1920 bright stage 
(E = 1 to 11) corresponds to an apparent period of 770 days, 
larger than the orbital one, while their subsequent decrease 
(E = 12 to 24) indicates a shorter period of 747 days. 

The same type of variability appeared again during the recent, 
1989 active phase. Observed changes in both the position and 
the shape of the minima are illustrated in the top right panel 
of Fig.~6 and in Fig.~2. During the transition {\em from the active 
phase to quiescence} ($A\rightarrow Q$ transitions), a systematic 
change in the minima positions at E = 49 to 51 corresponded 
to the apparent period of only 730 days. 
During the transition {\em from the quiescent to the active 
phase} ($Q\rightarrow A$ transitions) a significant change 
in the $O-C$ values by jump of +130 days was observed. 
The minima positions at E = 47 and E = 49 indicate an apparent 
period of 822 days. 
%
%
\begin{figure}[!t]
\centering
\begin{center}
\resizebox{15cm}{!}{\includegraphics[angle=-90]{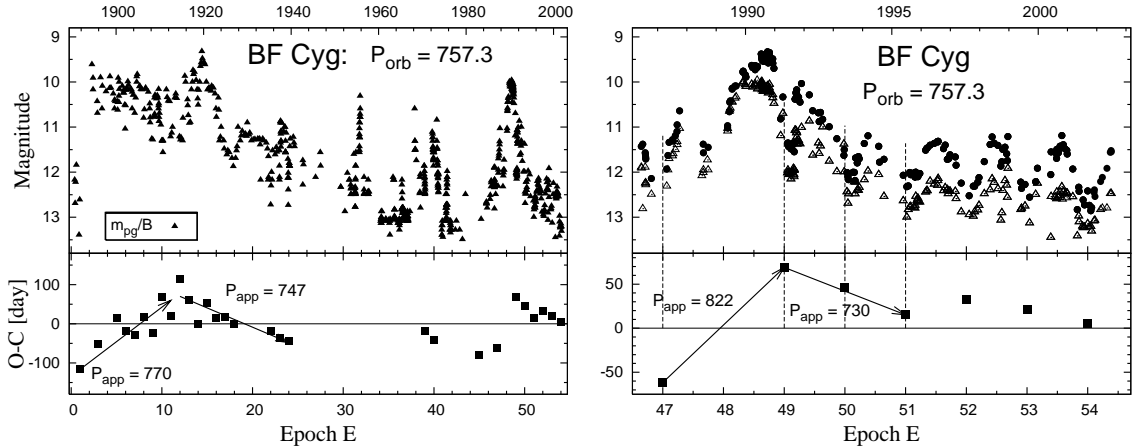}}
\caption[]{Historical and recent LCs of BF\,Cyg with 
the $O-C$ diagrams. During transition from quiescent 
to active phases and vice versa, the main source of radiation 
contributing to the optical continuum changes significantly 
its location and geometry in the symbiotic system. 
As a result the observed minima change their profile and 
position, what we indicate in the $O-C$ diagram as 
the apparent orbital changes. 
          }
\end{center}
\end{figure}
%
%

\subsection{Asymmetric shape of the H\,{\small II} zone}

To explain the observed apparent changes in the orbital period, 
the H\,{\small II} zone has to be extended asymmetrically 
with respect to the binary axis. Asymmetrical shape of 
the ionized region in symbiotic stars was also suggested, 
for example, by 

  (i) spectropolarimetric studies of \cite{schmid98} and 

  (ii) hydrodynamical calculations of the structure of 
stellar winds in symbiotic stars that include effects of 
the orbital motion \citep[e.g.][]{fw00}. 

\noindent
In both models the ionization front in the orbital plane 
is twisted, going from the side of the hot star that precedes 
its orbital motion, through the line joining the components, 
to the front of the cool star against its motion. 
Therefore, it is possible to assume that the main nebular 
region follows the S-shaped track from the front of the hot 
star orbital motion to the binary axis. 
Thus, the optically thick fraction of the H\,{\small II} 
region is prolonged so that its major axis at the orbital 
plane points the observer around the orbital phase 0.9. 

\subsection{Principle of apparent orbital changes}

During the $A\rightarrow Q$ transitions, the optically thick 
shell is gradually diluting, which leads to the increase of 
the hot star temperature and thus production of the ionizing 
photons. As a result, the optical continuum declines and changes 
significantly its nature -- from blackbody to nebular radiation 
(Fig.~3, bottom left and right panel). This process causes an
{\em expansion} of the H\,{\small II} zone and thus the change 
of the minimum profile from narrow one to broad wave 
throughout the orbital cycle. According to the asymmetry of 
the nebula (Sect.~6.2) the light minima occur prior to the 
time of spectroscopic conjunction. This behavior is also 
illustrated by the top panels of Fig.~3. Thus, during the 
$A\rightarrow Q$ transitions we indicate an apparent period, 
which is {\em shorter} than the orbital one. 

During $Q \rightarrow A$ transitions a sudden 
decrease in the luminosity of the ionizing photons results 
from rather rapid creation of the false relatively cool 
photosphere. This implies a disruption of the H\,{\small II} 
zone. Optical region is then (usually) dominated by the  
stellar radiation from the pseudophotosphere, and a narrow 
minimum (eclipse) is observed at the inferior conjunction 
of the giant. In such a case the time difference between 
the preceding broad minimum (at $\varphi \approx 0.9$) and 
the eclipse (at $\varphi \approx 0$) is 
$P_{\rm app} \approx P_{\rm orb} +0.1\times P_{\rm orb}$. 
This apparent change in the period happens suddenly -- the 
$O-C$ diagram indicates a jump in the residuals. 
In our illustration of BF\,Cyg (Fig.~6, minima just 
prior to the 1989 outburst), the timing of the broad 
minimum at the epoch E = 47 and the following eclipse at 
E = 49 corresponds to the apparent period 
$P_{\rm app} \approx 822$ days. 

\section{Concluding remarks}

\subsection{A complexity of LC profiles}

We have discussed only the fundamental variations in the LCs 
of symbiotic stars, i.e. those that can be explained with 
the aid of their basic model. However, LCs of symbiotic 
stars record a much larger variety of light changes that 
are unexpected and never repeat again. 
For example, during quiescence the wave-like variation is not 
a simple sinusoid, but alters its profile from cycle to cycle 
in both scales -- time and brightness 
\cite[e.g. EG\,And, AG\,Peg, AX\,Per,][]{sk+07}. 

During active phases the LC profiles are very heterogeneous. 
Eruptions arise unexpectedly with a rapid increase 
(e.g. RS\,Oph and most of AG\,Dra events), or more frequently 
with a gradual increase to the maximum within a few months 
\citep[e.g. recent outbursts of Z\,And and AG\,Dra,][]{sk+06,sk+07}. 
Also the recurrence time is an unpredictable phenomenon. 
For some objects no active phase has yet been recorded 
(e.g. SY\,Mus, RW\,Hya, EG\,And). For others, eruptions 
are scattered in historical LCs irregularly. For example, 
during 1994-98 period the AG\,Dra LC (Fig.~2) showed eruptions 
with a strict recurrence of $\sim$1 year 
\citep[e.g.][ Fig.~2 here]{viotti+07}. 
Previously, \cite{iijima87} suggested that since 1930 AG\,Dra 
periodically entered active stages with an interval of about 
15 years. However, from the beginning of its historical 
records of the brightness in 1890 to about 1927, it was quiet 
with a first strong outburst indicated around 1932 
\citep[][]{rob69}. 
Another illustrative example with this respect is YY\,Her 
\citep[][]{mu+97}.
In addition, an even more complex profile of the LC is observed 
when different types of eruptions (nova-like, Z\,And-type, 
flares) are superposed. An example here is the historical LC 
of BF\,Cyg \citep[Fig~6 and][ in detail]{sk+97}. 

\subsection{The problem of eclipses}

Eclipses can suddenly arise in the LC during active phases 
of symbiotics with a high orbital inclination (Sect.~4). 
However, their presence is not stable during each outburst 
of some symbiotic objects. For example, evidence for eclipses 
in the symbiotic triple system CH\,Cyg was reported by 
\cite{sk+96}. However, those produced by the inner symbiotic 
binary were observed only during the lower level of 
the activity (1967-71 and 1992-95). 

Generally, the depth of eclipses is very sensitive to the 
location of the main sources of radiation in the system 
and their relative contributions at the considered passband, 
which both can be subject to variation during different 
active phases. Therefore the eclipse effect can be observed 
only at specific brightness phases, at which the radiative 
contribution from a pseudophotosphere in the optical rivals 
that of the nebula. 
In addition, in the the case of CH\,Cyg, the effect of 
the precession of 
the inner orbit (i.e. that with the symbiotic pair) with 
a period of 6\,520 days and the precession cone opening 
angle of 35$^\circ$ \citep[][]{croc+02} is probably the main 
cause of the intriguing behavior of eclipses in this system. 

An additional problem connected with eclipses in symbiotic 
binaries is their width. In some cases it is too large 
to be explained by a simple eclipse of the hot object by 
the stellar disk of the giant. Examples here are BF\,Cyg 
\citep[][]{sk+97} and TX\,Cvn \citep[][]{sk+07}. 
In the case of Z\,And the broad eclipse profile suggested 
a disk-like structure for the hot object during active 
phases \citep[][]{sk03}. 

\subsection{Importance of the photometric monitoring}

The diversity of variations recorded in the LCs of symbiotic 
stars is thus far beyond our full understanding. 
The investigation of interactions between the cool giant and its 
hot luminous compact companion in a symbiotic binary requires 
simultaneous, multi-frequency observations from X-rays to radio 
wavelengths. This is an extremely challenging task, addressed 
mainly to large ground-based telescopes and those on 
satellites. 
In spite of this, the photometric monitoring of symbiotic stars, 
usually carried out with small telescopes, plays an important 
role in such research. I summarize some reasons as follows:
\begin{itemize}
\item
Monitoring usually first discovers an unpredictable sudden 
change in the brightness and thus can provide an alert for 
observation with other facilities. 
\item
Color indices can provide information about the nature 
of the composite continuum and thus to help to identify 
the responsible process. 
For example, the very negative intrinsic (i.e. dereddened 
and corrected for lines) $U-B$ index is usually connected 
with optical brightening that signals the energy 
conversion from the hot star to the nebular emission. 
Some examples were discussed by \cite{sk+06} and \cite{t+04}. 
\item
The eclipse profiles can help to recognize the structure of 
the hot active object -- a spherical or a disk-like structure. 
Disentangling the color indices during the totality allow 
us to quantify contribution from the non-eclipsed fraction 
of the nebula (Sect.~4). 
\item
The minima during quiescence, whose profiles reflect the 
geometry of the nebula, can determine the difference between 
the simplified ionization structure (cf. Fig.~4) and the real 
situation including effects of the binary motion and accretion. 
\item
Variation in the LC profile in the $UB[V]$ bands around the 
orbital phase $\varphi \sim 0.5$ (e.g. presence/absence of 
a secondary minimum) can probe the extension of 
the nebula -- if it is closed or open in the sense of 
the STB model (Fig.~4, Sect.~5.2). 
\item
A double-wave profile in the $[V]RI$ passbands implies the 
possibility of the ellipsoidal shape of the red giant due 
to tidal distortion and thus can discriminate the type of 
mass transfer process (via the wind or the Roche lobe 
overflow?). 
\item
An intrinsic variability of the giant component in symbiotic 
binaries monitored in the $VRI$ passbands can provide 
physical parameters for such 'pulsation-type' of variability. 
Objects as CH\,Cyg, CI\,Cyg, AG\,Peg, AR\,Pav are promising 
candidates here \citep[][]{mmk92,b+p91,sk+07,sk+00}. 
\item
Multicolor LCs, if properly corrected for influence of 
emission lines \citep[][]{sk07}, can provide a satisfactory 
tool to calibrate spectroscopic observations and thus to be 
useful in determining other physical parameters. 
\item
Evolution in the LC profile at the very beginning stage 
of outbursts is of particular importance to mapping 
the process igniting the eruption \citep[e.g.][]{sok+06}. 
\end{itemize} 

\vspace*{10mm}

\noindent
{\em Acknowledgments.}
The author thanks the anonymous referee for useful comments 
and is also grateful to Michael Saladyga for the help with 
editing the original text. 
This work was supported by the Slovak Academy of Sciences
grant No. 2/7010/7.

\end{document}